\documentclass[aps,pre,twocolumn,superscriptaddress,showpacs,notitlepage]{revtex4-1}

\usepackage{amsmath}
\usepackage{amssymb}
\usepackage[all]{xy}
\input xy
\xyoption{matrix}
\xyoption{2cell}
\xyoption{arrow}
\xyoption{curve}
\xyoption{all}

\usepackage{hyperref}
\usepackage{amsmath}
\usepackage{graphicx}
\usepackage[latin1]{inputenc}
\usepackage{ulem}
\usepackage{epstopdf}

\begin{document}

\title{Approximating Metal-Insulator Transitions}
 \author{Carlo Danieli}
\affiliation{New Zealand Institute for Advanced
  Study, Centre for Theoretical Chemistry and Physics, Massey
  University, Auckland, New Zealand}
\author{Kristian Rayanov}
\affiliation{New Zealand Institute for Advanced
  Study, Centre for Theoretical Chemistry and Physics, Massey
  University, Auckland, New Zealand}
\author{Boris Pavlov}
\affiliation{New Zealand Institute for Advanced
  Study, Massey
  University, Auckland, New Zealand}
\author{Gaven Martin}
\affiliation{New Zealand Institute for Advanced
  Study, Massey
  University, Auckland, New Zealand}
\author{Sergej Flach}
\affiliation{New Zealand Institute for Advanced
  Study, Centre for Theoretical Chemistry and Physics, Massey
  University, Auckland, New Zealand}
\date{\today}

\begin{abstract}
We consider quantum wave propagation in one-dimensional quasiperiodic lattices.
We propose an iterative construction  of quasiperiodic potentials from sequences of potentials with increasing spatial period. 
At each finite iteration step the eigenstates reflect the properties of the limiting quasiperiodic potential properties up to a controlled maximum system size.
We then observe approximate metal-insulator transitions (MIT) at the finite iteration steps. We also report evidence on mobility edges which are at variance to the celebrated Aubry-Andre model. The dynamics near the MIT shows a critical slowing down of the ballistic group velocity in the metallic phase similar to the divergence of the localization length in the insulating phase. 
\end{abstract}

\maketitle

\section{Introduction}

Wave localization in random potentials has been intensively studied ever since its prediction by Anderson in 1958 \cite{anderson1958}. 
Notably uncorrelated random potentials in space dimension $d=1$ will lead to complete localization. Metal-Insulator transitions (MIT) and possible mobility edges (energies separating
delocalized (metallic) from localized (insulating) eigenstates) will typically appear in systems with space dimension $d=3$. Later studies of potentials with correlated disorder
have shown that the dimension restriction for an MIT can be lowered to $d=1$ if there are sufficiently strong correlations in the disorder potential \cite{correlateddisorder}.

It came as a surprise that the quasiperiodic potential introduced by Aubry and Andr\'e (AA) in 1980 allowed an MIT when $d=1$ \cite{AA}.
This MIT is tuned by the strength of a $\cos(2\pi \alpha \ell)$ potential whose period $1/\alpha$ is irrational and therefore incommensurate with the lattice spacing $\Delta  \ell=1$.
Analytical results were obtained thanks to a duality principle relating states and spectra in direct and Fourier space. This very principle however prevents the appearance
of mobility edges (the MIT of the AA model is not dependent on the eigenenergy). Attempts to generalize these results to other quasiperiodic potentials for $d=1$ showed that the  localized regime could be maintained irrespective of the potential strength \cite{grempel82}.  Further,  Fibonacci sequence based potentials kept the system at the critical point \cite{kohmoto83,ostlund83},
and a mix of two Aubry-Andr{\'e} potentials with different periods allowed  the appearance of a mobility edge \cite{hiramoto89}.
A recent study of two interacting particles in the Aubry-Andr\'e model established the appearance of metallic correlated bound states deep in the insulating regime for a single particle \cite{Flach12}.

Experimental studies on light propagating through optical waveguide networks \cite{lahini09} and ultracold atomic clouds expanding in optical potentials \cite{roati08} successfully
tested the MIT within the Aubry-Andr\'e model. The flexibility in the choice of potentials within these studies makes them ideal testing grounds for other quasiperiodic potentials.
At the same time these systems have finite size, and have unavoidable precision limitations on the generated potentials \cite{larcher09}. Desired effects like the MIT or mobility edges are therefore
needed only up to that precision, and only to be observable on these length scales. 

In this work we present a systematic and constructive way to approximate a quasiperiodic potential
by a periodic one. Each approximation is defined both by its period and  by the convergence criteria of the amplitude sequence of higher harmonics. This flexibility
allows us to obtain a wide variety of quasiperiodic potentials which can be expected to exhibit the above phenomena. In addition the experimentally relevant length scale can be easily
taken into account by the corresponding periodic approximation of a quasiperiodic potential. 

The paper is structured as follows: in the next section we briefly discuss the Aubry-Andr\'e model and some less known properties of wave packet spreading.
In Section III we introduce the construction principle for the new class of quasiperiodic pontentials. In section IV we discuss the main properties of localized and extended states for particular
amplitude sequences. Finally we conclude and summarize.

\section{Aubry-Andr\'e model}

Consider the  $d=1$ dimensional discrete Schr\"odinger operator $H:\ell^2(\mathbb{Z})\longrightarrow\ell^2(\mathbb{Z})$, also known as Aubry-Andr\'e model, defined by
\begin{equation}
\label{eq:H}
(H\psi)_l := \epsilon_l\psi_l + \psi_{l+1} + \psi_{l-1}\ ,\quad l\in\mathbb{Z}\ ;
\end{equation}
with quasi-periodic potential 
\begin{equation}
\label{eq:H2}
\epsilon_l = \lambda\cos(2\pi(\alpha l+\beta))\ ,\quad \alpha\in\mathbb{R}\setminus\mathbb{Q}\ ,
\end{equation}
with a positive real strength $\lambda>0$, $\beta\in\mathbb{R}$, and irrational $\alpha$. 
Due to the self-dual character at $\lambda=2$, the model exhibits a transition between a metallic phase for $\lambda\in]0,2[$ and an insulating phase when $\lambda\in]2,+\infty[$.
It is also well-known \cite{AA} that, for any given $\lambda$ in the insulating phase, all normal modes decay exponentially in space as $e^{-l/\xi}$, with the localisation length $\xi = 1 / \ln (\lambda/2)$ being independent of the eigenenergy.

\begin{figure}[!h]
 \centering
 \includegraphics[width=.41\textwidth]{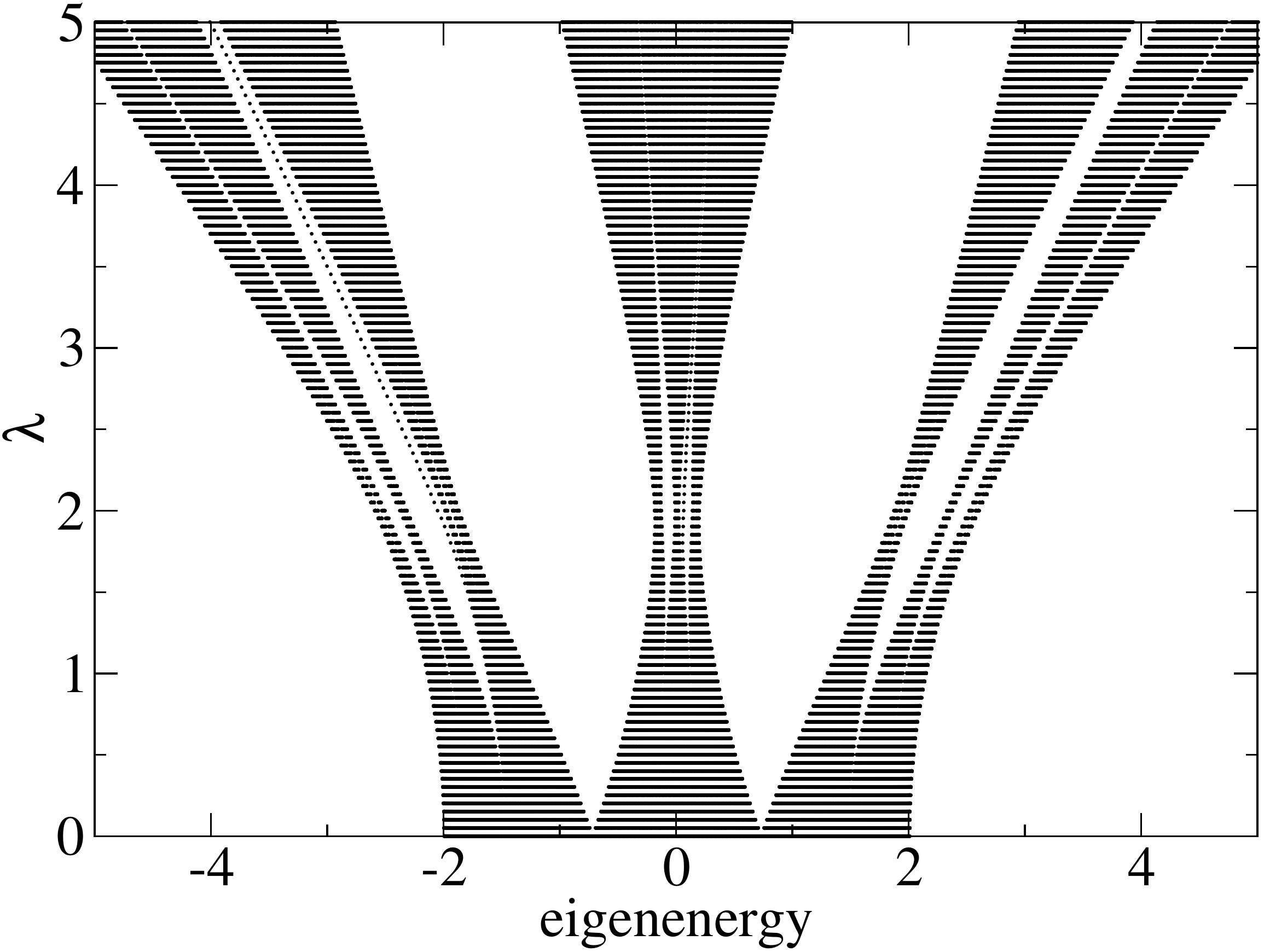}
 \caption{Spectrum $\sigma_\lambda$ of the Aubry-Andr\'e model when  $\alpha=(\sqrt{5} - 1)/2$. }
 \label{fig1}
\end{figure}
The eigenenergy spectrum $\sigma_\lambda(\alpha,\beta)$ (eg. Fig\ref{fig1}) of the Aubry-Andr\'e model is a fractal set with a self-similiar structure which does not depend
on $\beta$.  When $\alpha=(\sqrt{5} - 1)/2$ is the golden mean, it consists of three bands, which again consist of three sub-bands and so forth (Fig. \ref{fig1}). The spectrum has a  Cantor set structure for all $\lambda\neq 0$ \cite{Av1}, and its Lebesgue measure \cite{Av2} can be found analytically by the formula
\[
\mu(\sigma_\lambda) = 2|2-|\lambda||\ .
\]
For $\lambda\in]0,2[$ (metallic phase), the model has absolutely continuous spectrum \cite{Avilla},  while for   $\lambda\in]2,+\infty[$ (insulating phase) it is a  purely point spectrum \cite{Avilla3}. At the critical value $\lambda=2$, the spectrum   is   purely singular and continuous,   \cite{Avilla2}. The spectrum's fractal dimension is $1$ for $\lambda\neq 2$, and $0.5$  for $\lambda=2$ \cite{Tang}.

Remarkably, there are much  more exact results for the  spectral properties of the Aubry-Andr\'e model  as compared to the dynamical properties of wave packet spreading 
\begin{equation}
i\dot{\psi}_l = (H\psi)_l
\end{equation}
where only a few
numerical results are known \cite{hiramoto88,larcher09}.

\medskip

Consider a single site  excitation for a normalized wave function $\sum_l |\psi_l|^2 = 1$. Its evolution in time leads to a time dependent distribution $n_l(t) = |\psi_l(t)|^2$.
The dynamics can be characterized by the second moment $m_2 = \sum_l ( l - \sum_k kn_k )^2 n_l$.
 We can distinguish different spreading regimes \cite{hiramoto88,larcher09}:
\begin{displaymath}
m_2 \sim  \left\{ \begin{array}{ll}
v_g^2\ t^2 & \ \ \textrm{Ballistic Spreading}\\
D t & \ \ \textrm{Diffusive Spreading} \qquad\qquad\ ,\\
\xi^2 & \ \ \textrm{Localisation}
\end{array} \right.
\end{displaymath}
where the coefficients are: the group velocity $v_g$, the diffusion coefficient $D$ and the localisation length $\xi$. 
In particular,  spreading is ballistic in the metallic regime $ \lambda < 2$ and diffusive at the critical point $\lambda=2$ \cite{hiramoto88,larcher09}.

\begin{figure}[!h]
 \centering
 \includegraphics[width=.44\textwidth]{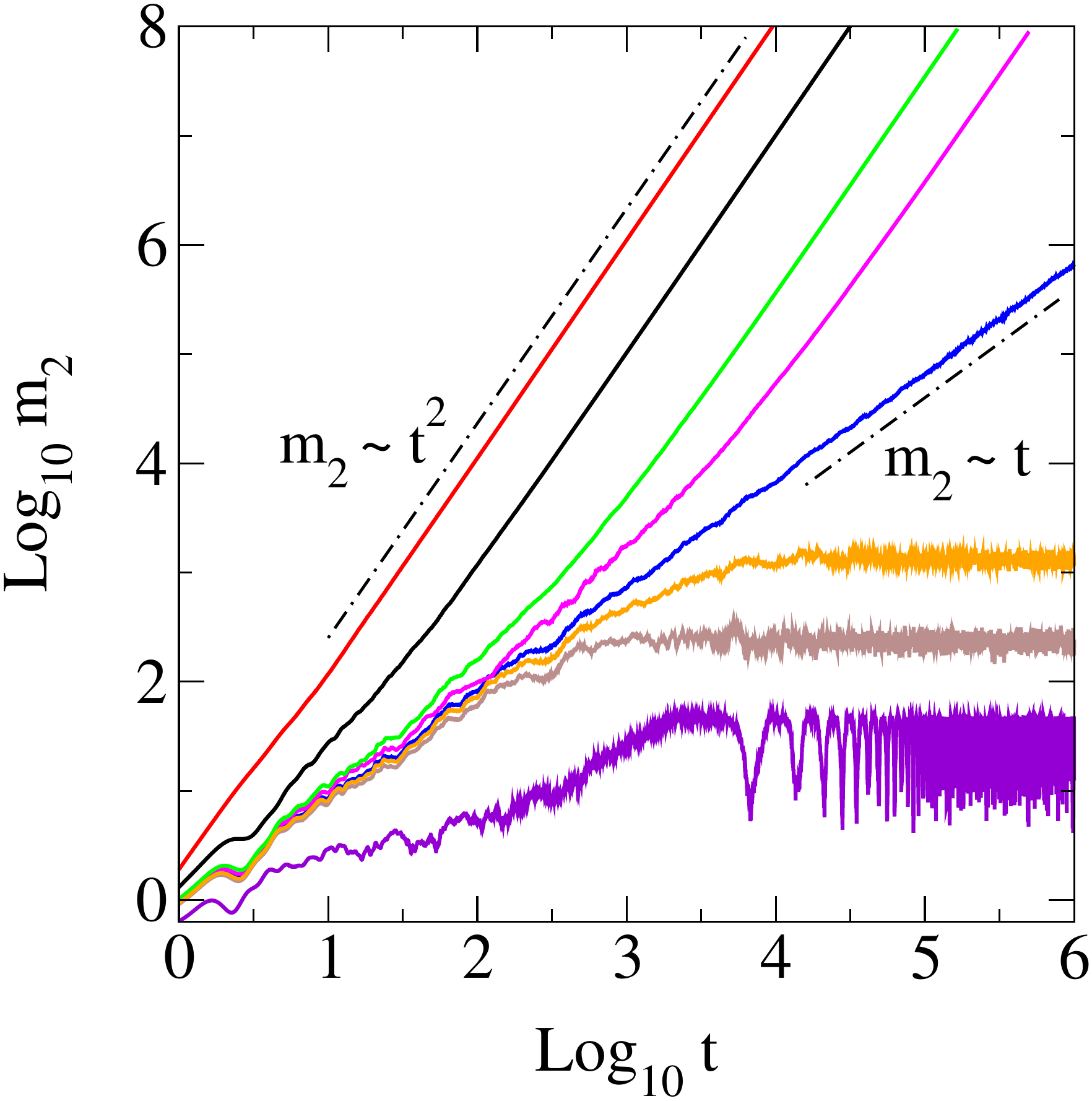}
 \caption{The second moment $m_2$ for a single site excitation as a function of time in a log-log plot. From top to bottom: $\lambda=0.5$ (red), $\lambda=1.5$ (black), $\lambda=1.9$ (green), $\lambda=1.97$ (orange, )
$\lambda=2$ (blue), $\lambda=2.05$ (brown) and $\lambda=2.5$ (violet). The dashed-dotted lines indicate power laws $m_2 \sim t$ (diffusive) and $m_s \sim t^2$ (ballistic).
Here $\alpha=(\sqrt{5} - 1)/2$ and $\beta=0$.
}
 \label{fig2}
\end{figure}

 We plot the time evolution of the second moment in Fig.\ref{fig2}.
We see that, indeed, for $\lambda=2$ spreading appears to be diffusive, and we can confirm the predicted ballistic asymptotics  in the metallic regime and localisation in the insulating regime.
However, we also observe a diffusive transient in these regimes, which becomes longer the closer one gets to the critical point. 
\begin{figure}[!h]
 \centering
 \includegraphics[width=.44\textwidth]{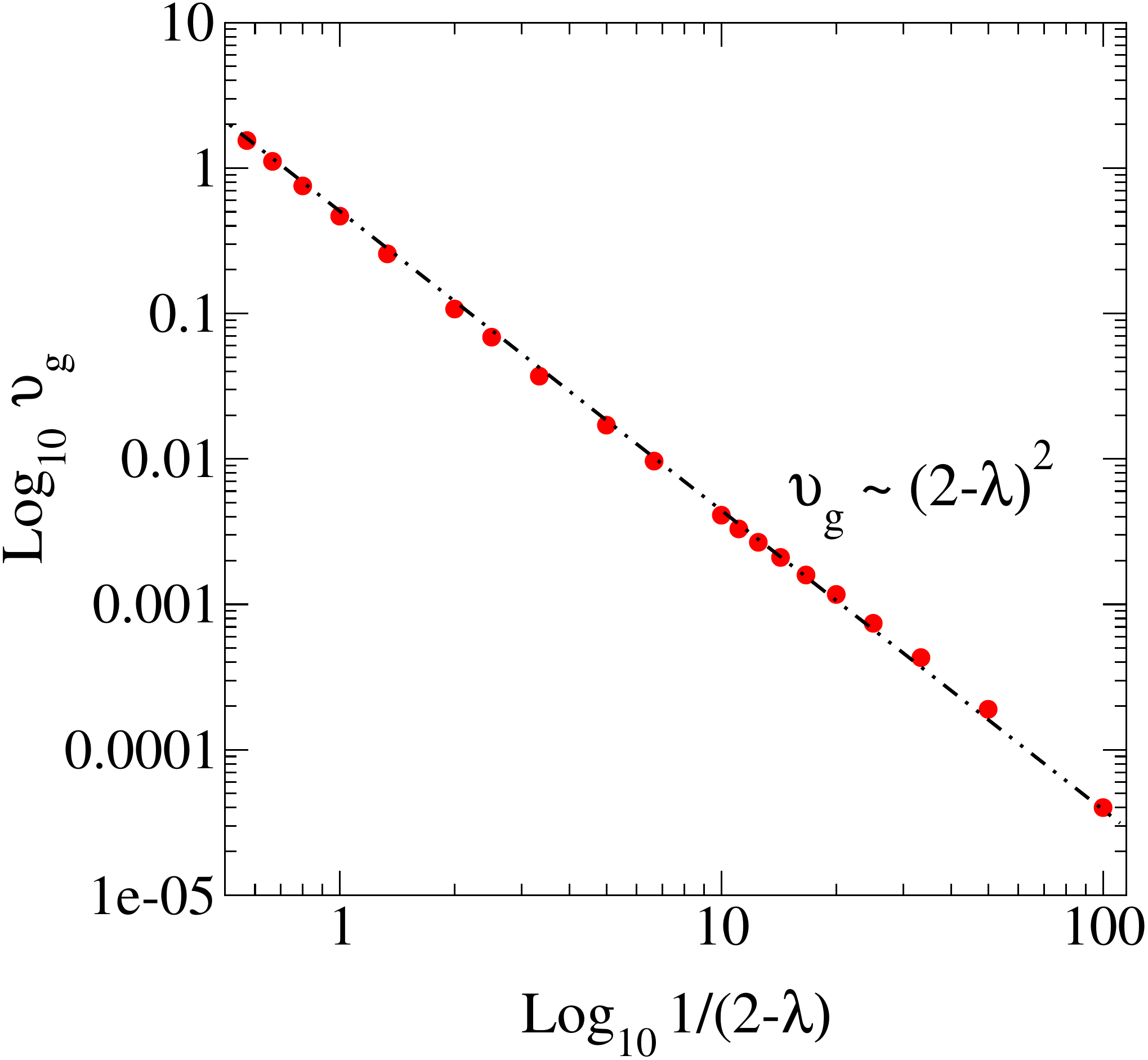}
 \caption{The dependence of the group velocity $v_g$ on $1/(2-\lambda)$ in the metallic phase in a log-log plot.
 The symbols are the actual computed values, the dashed line corresponds to the law $v_g \sim (2-\lambda)^2$.}
 \label{fig3}
\end{figure}
We compute the group velocity $v_g$ which is given by the square root of $m_2$ up to a constant prefactor, and plot it in Fig.\ref{fig3}. We find that it vanishes at the critical point as $v_g \sim (2-\lambda)^2$.

It remains an intriguing task to explain these spreading regimes in their relation to the Cantor spectrum of the model.
This concerns in particular the crossover time $\tau_{\lambda}$, the critical exponent for $v_g$, and the question why spreading into a large but finite localization volume in the
insulating regime happens to be diffusive and not ballistic, as for one-dimensional uncorrelated disorder (e.g.  \cite{rayanov13}).
These observations may   be rather special features of the highly symmetric Aubry-Andr\'e model which enjoys duality. 

\section{Cantor-like constructions of a class of quasiperiodic potentials} 

We now  construct quasiperiodic potentials in a systematic way approximating them by periodic potentials at each iteration step. The standard construction  
of a $1/3$ Cantor set (cut out the middle third of an interval, then repeat with the reamining subintervals ad infinitum) gives
a set of measure zero and nonzero Hausdorff dimension $\frac{\log2}{\log{3}}$. Here we modify this procedure for the eigenenergy spectrum effectively by stretching and changing the
cutting ratio. This is all achieved by defining sequences of periodic potentials with increasing spatial period.
For simplicity we shall start with an example which has a three-band structure (where each band again subdivides into three subbands etc) similar to the golden mean case of 
the Aubry-Andr\'e spectrum before we generalize. 

\subsection{The scheme }

Consider a sequence of discrete periodic functions $E_k(l)$ with $k=1,2,3,...$. Each function is periodic with spatial period $L_k=3^k$: $E_k(l+L_k) = E_k(l)$.
The function $E_1$ is defined for $m\in\mathbb{Z} $ by
\begin{displaymath}
E_1(l) := \left\{ \begin{array}{ll}
-a_1 \quad & \textrm{$l=3 m$}\\
0  \quad & \textrm{$l=3 m + 1$} \\
+a_1 \quad & \textrm{$l=3 m + 2$}\\
\end{array} \right. 
\end{displaymath}
with the following schematic form:
\[
\xy
(12,0)*{} ; (83,0)*{} **\dir{-} ; (83,0)*{\dir{>}} ; (20,-3.5 )*{\bullet} ;  (25,0)*{\bullet} ;  (30,3.5)*{\bullet}  ; (35,-3.5)*{\bullet} ;  (40,0)*{\bullet} ;  (45,3.5)*{\bullet} ; (50,-3.5)*{\bullet} ;  (55,0)*{\bullet} ;  (60,3.5)*{\bullet} ; (65,-3.5)*{\bullet} ;  (70,0)*{\bullet} ;  (75,3.5)*{\bullet} ;  (20,-1)*{} ;  (20,1)*{} **\dir{-} ;  (25,-1)*{} ;  (25,1)*{} **\dir{-} ; (30,-1)*{} ;  (30,1)*{} **\dir{-} ; (35,-1)*{} ;  (35,1)*{} **\dir{-} ; (40,-1)*{} ;  (40,1)*{} **\dir{-} ; (45,-1)*{} ;  (45,1)*{} **\dir{-} ; (50,-1)*{} ;  (50,1)*{} **\dir{-} ; (55,-1)*{} ;  (55,1)*{} **\dir{-} ; (60,-1)*{} ;  (60,1)*{} **\dir{-} ; (65,-1)*{} ;  (65,1)*{} **\dir{-} ; (70,-1)*{} ;  (70,1)*{} **\dir{-} ; (75,-1)*{} ;  (75,1)*{} **\dir{-} ; (18,3.5)*{} ; (77,3.5)*{} **\dir{.} ; (15.5,3.5)*{a_1} ; (18,-3.5)*{} ; (77,-3.5)*{} **\dir{.} ; (15,-3.5)*{-a_1} ; (84.5,-3.5)*{l\in\mathbb{Z}} ; 
\endxy
\]
The second function $E_2$ has period $L_2=9$:
\begin{displaymath}
E_2(l) := \left\{ \begin{array}{ll}
-a_2 \quad & \textrm{$l = 9m+\{0,1,2\}$}\\
0  \quad & \textrm{$l = 9m+\{3,4,5\}$} \qquad ,\quad m\in\mathbb{Z} \\
+a_2 \quad & \textrm{$l = 9 m+\{6,7,8\}$}\\
\end{array} \right.
\end{displaymath}
and the schematic form
\[
\xy
(12,0)*{} ; (83,0)*{} **\dir{-} ; (83,0)*{\dir{>}} ; (20,-3)*{\bullet} ;  (25,-3)*{\bullet} ;  (30,-3)*{\bullet}  ; (35,0)*{\bullet} ;  (40,0)*{\bullet} ;  (45,0)*{\bullet} ; (50,3)*{\bullet} ;  (55,3)*{\bullet} ;  (60,3)*{\bullet} ; (65,-3)*{\bullet} ;  (70,-3)*{\bullet} ;  (75,-3)*{\bullet} ;  (20,-1)*{} ;  (20,1)*{} **\dir{-} ;  (25,-1)*{} ;  (25,1)*{} **\dir{-} ; (30,-1)*{} ;  (30,1)*{} **\dir{-} ; (35,-1)*{} ;  (35,1)*{} **\dir{-} ; (40,-1)*{} ;  (40,1)*{} **\dir{-} ; (45,-1)*{} ;  (45,1)*{} **\dir{-} ; (50,-1)*{} ;  (50,1)*{} **\dir{-} ; (55,-1)*{} ;  (55,1)*{} **\dir{-} ; (60,-1)*{} ;  (60,1)*{} **\dir{-} ; (65,-1)*{} ;  (65,1)*{} **\dir{-} ; (70,-1)*{} ;  (70,1)*{} **\dir{-} ; (75,-1)*{} ;  (75,1)*{} **\dir{-} ; (18,3)*{} ; (77,3)*{} **\dir{.} ; (15.5,3)*{a_2} ; (18,-3)*{} ; (77,-3)*{} **\dir{.} ; (15,-3)*{-a_2} ; (84.5,-3.5)*{l\in\mathbb{Z}} ; 
\endxy
\]
Higher order functions $E_k(l)$ are defined in a similar way and are characterized by the corresponding amplitude $a_k$:
\begin{displaymath}
E_k(l) := \left\{ \begin{array}{ll}
-a_k & \textrm{if $l=3^km+q$}\\
0 & \textrm{if $l=3^{k-1}(3m+1)+q$} \qquad , \quad m,q\in\mathbb{Z}\\
+a_k & \textrm{if $l=3^{k-1}(3m+2)+q$}
\end{array} \right.
\end{displaymath}
with $0 \leq q \leq 3^{k-1} -1$, and their schematic form is
\[
\xy
(5,0)*{} ; (87,0)*{} **\dir{-} ;  (87,0)*{\dir{>}} ; (88.5,-3.5)*{l\in\mathbb{Z}}  ; (10,-1)*{} ;  (10,1)*{} **\dir{-} ; (13,-.7)*{} ;  (13,.7)*{} **\dir{-} ; (27.3,-.7)*{} ;  (27.3,.7)*{} **\dir{-} ; (30.3,-.7)*{} ;  (30.3,.7)*{} **\dir{-} ;(33.3,-1)*{} ;  (33.3,1)*{} **\dir{-} ; (36.3,-.7)*{} ;  (36.3,.7)*{} **\dir{-} ; (50.6,-.7)*{} ;  (50.6,.7)*{} **\dir{-} ; (53.6,-.7)*{} ;  (53.6,.7)*{} **\dir{-} ; (56.6,-1)*{} ;  (56.6,1)*{} **\dir{-} ; (59.6,-.7)*{} ;  (59.6,.7)*{} **\dir{-} ; (74,-.7)*{} ;  (74,.7)*{} **\dir{-} ; (77,-.7)*{} ;  (77,.7)*{} **\dir{-} ;  (80,-1)*{} ;  (80,1)*{} **\dir{-} ; 
(10,-6)*{0} ; (35.3,-6)*{3^{k-1}} ; (59.6,-6)*{2\cdot 3^{k-1}} ; (80,-6)*{3^k} ;
  (10,-3)*{\bullet} ; (13,-3)*{\bullet} ; (27.3,-3)*{\bullet} ; (30.3,-3)*{\bullet}  ; (33.3,0)*{\bullet} ; (36.3,0)*{\bullet} ; (50.3,0)*{\bullet} ; (53.3,0)*{\bullet} ; (56.3,3)*{\bullet} ; (59.6,3)*{\bullet} ; (74,3)*{\bullet} ; (77,3)*{\bullet} ; (7,3)*{} ; (83,3)*{} **\dir{.} ; (4,3)*{a_k} ; (7,-3)*{} ; (83,-3)*{} **\dir{.} ; (3.5,-3)*{-a_k} ; (10,-3)*{} ; (30.3,-3)*{} **\dir{--} ; (56.3,3)*{} ; (77,3)*{} **\dir{--} ; 
\endxy
\]

Now we consider the superposition of all $E_k(l)$ for all $1 \leq k \leq K$:
\begin{equation}
\label{eq:NP1}
\epsilon_{l,K}:=\lambda \sum_{k=1}^K E_k(l)\ ,\quad \forall l\in\mathbb{Z}\ ,
\end{equation}
where $\lambda>  0$ is again the potential strength parameter. The potential $\epsilon_{l,K}$ is periodic with period $L_K$. 
For example, for $K=2$ we obtain a potential of period $L_2=9$: $\epsilon_{l,2} = E_1(l)+E_2(l)$ with the spatial profile
\[
\xy
(12,0)*{} ; (83,0)*{} **\dir{-} ; (83,0)*{\dir{>}} ; (20,-6.5)*{\bullet} ;  (25,-2)*{\bullet} ;  (30,1.5)*{\bullet}  ; (35,-3.5)*{\bullet} ;  (40,0)*{\bullet} ;  (45,3.5)*{\bullet} ; (50,-1.5)*{\bullet} ;  (55,2)*{\bullet} ;  (60,6.5)*{\bullet} ; (65,-6.5)*{\bullet} ;  (70,-2)*{\bullet} ;  (75,1.5)*{\bullet} ;  (20,-1)*{} ;  (20,1)*{} **\dir{-} ;  (25,-1)*{} ;  (25,1)*{} **\dir{-} ; (30,-1)*{} ;  (30,1)*{} **\dir{-} ; (35,-1)*{} ;  (35,1)*{} **\dir{-} ; (40,-1)*{} ;  (40,1)*{} **\dir{-} ; (45,-1)*{} ;  (45,1)*{} **\dir{-} ; (50,-1)*{} ;  (50,1)*{} **\dir{-} ; (55,-1)*{} ;  (55,1)*{} **\dir{-} ; (60,-1)*{} ;  (60,1)*{} **\dir{-} ; (65,-1)*{} ;  (65,1)*{} **\dir{-} ; (70,-1)*{} ;  (70,1)*{} **\dir{-} ; (75,-1)*{} ;  (75,1)*{} **\dir{-} ; (84.5,-3.5)*{l\in\mathbb{Z}} ; 
\endxy
\]

The same construction could be done for sequences of periodic functions $E_k(l)$ with $k=1,2,3,...$ of spatial period $L_k=s^k$ with $s\in\mathbb{N}$: $E_k(l+L_k) = E_k(l)$. We can simplify notations by using the definition $[l]_m\equiv l \; mod \; m$ to arrive at
\[
E_k(l)= \Bigg( \bigg\lfloor\frac{[l]_{s^k}}{s^{k-1}} \bigg\rfloor - \delta(s) \Bigg) a_k =: \phi_k(l) a_k\ ,\quad\forall l\in\mathbb{Z}\ .
\]
where $\bigg\lfloor . \bigg\rfloor$ denotes taking the integer part and
\[
\delta(s) := \left\{
\begin{array}{rl} 
\big\lfloor s / 2 \big\rfloor & \text{if } $s$\ \text{odd},\\
(s-1)/2 & \text{if } $s$\ \text{even}.
\end{array} \right.
\]
The final expression for the potential $\epsilon_{l,K}$ is 
\begin{equation}
\label{eq:pot1}
\epsilon_{l,K}:=\lambda \sum_{k=1}^K \phi_k(l) a_k\ ,\quad \forall l\in\mathbb{Z}\ , 
\end{equation}
where $\{a_k\}_{k=1}^K$ is the generating sequence of the potential and $\{\phi_k(l)\}_{k=1}^K$ the partitioning sequence. 

In the absence of any potential $\epsilon$ the spectrum of the operator (\ref{eq:H}) is given by one band $\sigma=2\cos p$ where $p$ is a Bloch wave number.
For $K=1$ the spectrum splits into three bands which are separated by two gaps. The first step in the above construction  therefore cuts two segments out of
the one band spectrum. At the next step of the approximation $K=2$, the spatial period $L_2=9$, and the spectrum consists now of 9 bands and 8 gaps.
Therefore each of the three bands of the $K=1$ spectrum is split into three narrower ones, with the new subgaps removing parts of the $K=1$ bands.
At the same time the bands edges may also shift, thus we obtain a Cantor-like iterative construction.

\subsection{The quasiperiodic limit}

We now push the iterative construction to its limit by extending (\ref{eq:pot1}) to an infinite sum
\begin{equation}
\label{eq:NPinf}
\epsilon_{l}:=\lambda \sum_{k=1}^{+\infty} \phi_k(l) a_k\ ,\hskip10pt  l\in\mathbb{Z}\ .
\end{equation}
The finiteness of this sum depends on the convergence properties of the generating sequence.   We make the following definition:  a sequence $\{\epsilon_l\}_{ l\in\mathbb{Z}}$ is {\it quasiperiodic}  if for every $\delta>0$,  there is $T=T(\delta) >0$ such that for all $l\in \mathbb{Z}$ we have $|\epsilon_{l+T}-\epsilon_{l}|<\delta$.  We have the following:

{\bf Lemma.} Let's consider a family $\{ \{\epsilon_l^k\}_{ l\in\mathbb{Z}}\}_{k=0}^{+\infty}$ of potentials, and suppose that for each $k$ the sequence $\{\epsilon^{k}_{l}\}_{l\in\mathbb{Z} }$ is periodic with period $L_k$ and the tails satisfy
\[ 
\lim_{N\to\infty} \sup_{l\in\mathbb{Z}} \sum_{k=N}^{\infty} \epsilon_{l}^{k} = 0\ .
\]
Then the sequence $\{\tilde{\epsilon}_{l}\}_{l\in\mathbb{Z} }$ defined by the sum
\begin{equation}
\tilde{\epsilon}_{l} = \sum_{k=0}^{\infty} \epsilon_{l}^{k}
\end{equation}
is quasiperiodic.

The proof is simply to observe that  any partial sum $\sum_{k=0}^{N} \epsilon_{l}^{k}$ is periodic with period $T_N:={\it lcm}(L_1,L_2,\ldots,L_N)$ 
(here $lcm$ means least common multiple)
and that the tail sums $ \sum_{k=N}^{\infty} \epsilon_{l}^{k}$ are uniformly small,  independent of $l$. So, for a given $\delta>0$, we choose $N=N(\delta)$ such that 
\[
\sup_{l\in\mathbb{Z}}  \sum_{k=N+1}^{\infty} \epsilon_{l}^{k} <\frac{\delta}{2}\ .
\]
It follows that $\forall l\in\mathbb{Z}$ we have 
\[
\begin{split}
|\tilde{\epsilon}_{l+T_N}-\tilde{\epsilon}_{l}| &= \Big|\sum_{k=0}^{N} \epsilon_{l + T_N}^{k} + \sum_{k=N+1}^{+\infty} \epsilon_{l + T_N}^{k}  \\
&- \quad\sum_{k=0}^{N} \epsilon_{l}^{k} \quad- \sum_{k=N+1}^{+\infty} \epsilon_{l}^{k} \Big| < 2\cdot \frac{\delta}{2} = \delta\ .
\end{split}
\]

As a direct corollary we see that the potential defined by (\ref{eq:NPinf}) is quasiperiodic. 

Thus the above class of quasiperiodic potentials,  which is defined by its generating sequence $\{ a_k \}_{k=1}^{+\infty}$  and the choice of the integer $s$ of the partitioning sequence $\{\phi_k(l)\}_{k=1}^{+\infty}$, is quasiperiodic and  has a self-similar spectrum.  

\section{Numerical results}

In this section, we first analyze a particular case A of potentials, defined by the geometric sequence $a_k=\mu^k$ (Model A), for a real value $\mu\in]0,1[$ and $s=3$
\begin{equation}
\epsilon_{l,K}:=\lambda\sum_{k=1}^K \phi_k(l) \mu^k\ ,\quad \forall l\in\mathbb{Z}\ .
\label{eq:modelA}
\end{equation}
First we consider $\lambda=1$. 
\begin{figure}[!h]
  { \includegraphics[ width=.85\columnwidth]{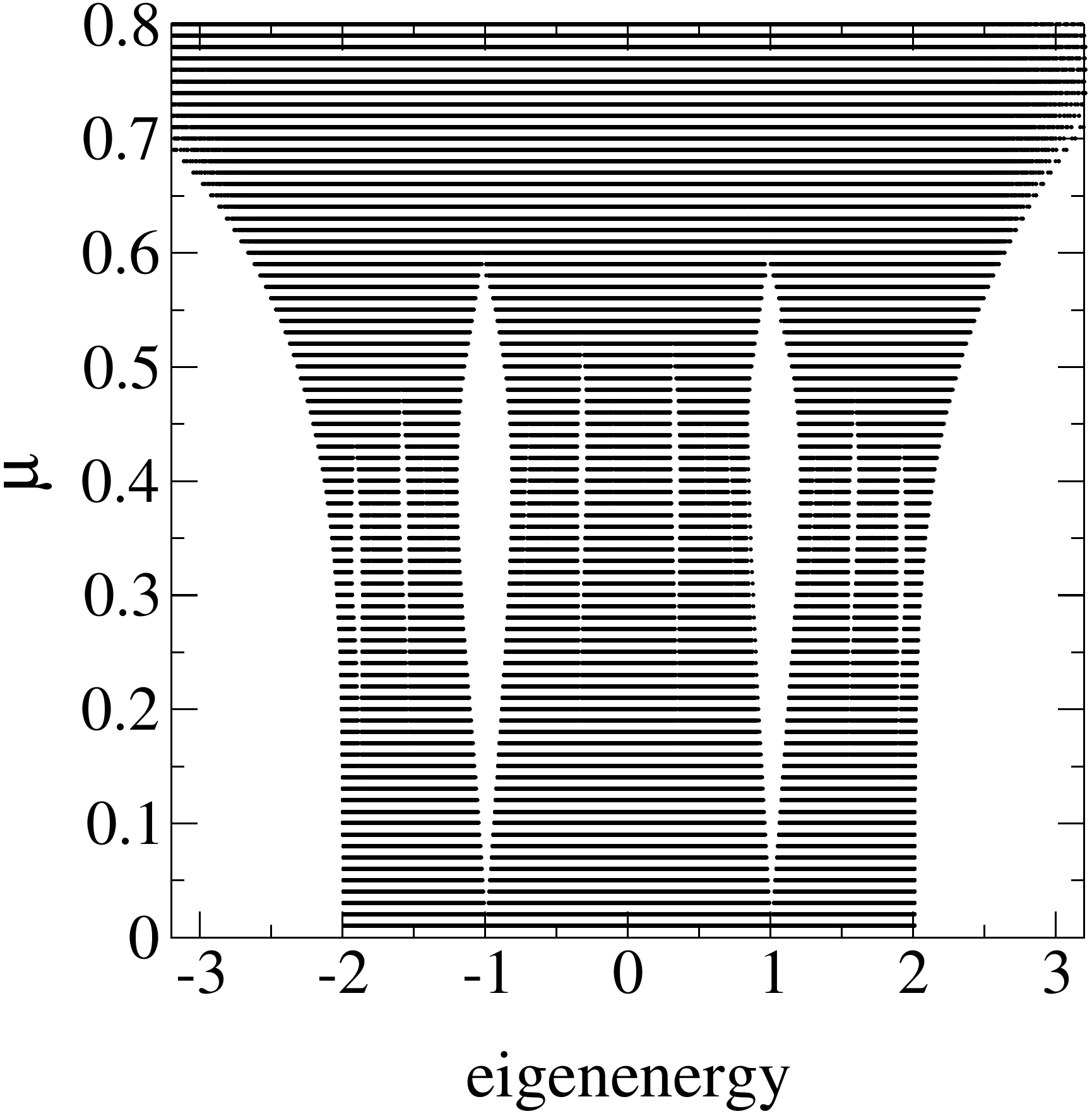}}
 \caption{Model A. The eigenenergy spectrum versus $\mu$. Here $s=3$, $\lambda=1$ and $K=9$. }
\label{fig:fig4}
\end{figure}

In Fig. \ref{fig:fig4} we show how the spectrum changes for different potential strength parameter $\mu\in]0,1[$. For $\mu$ from $0$ to $0.5$ the width of the spectrum and its sub-band shrinks and then, for $\mu\geq 0.5$ it starts to stretch. A similar effect is seen in the Aubry-Andr\'e model spectrum around the transition value $\lambda=2$. \\

To characterize localization of the corresponding eigenstates, we compute the Participation Number $P = 1 / \sum_l |\psi_l|^4$ of each eigenmode $(\psi_l)_{l\in\mathbb{Z}}$ and consider the maximum $P_{max}$ for a given $\mu$.
\begin{figure}[!h]
\includegraphics[width=.85\columnwidth]{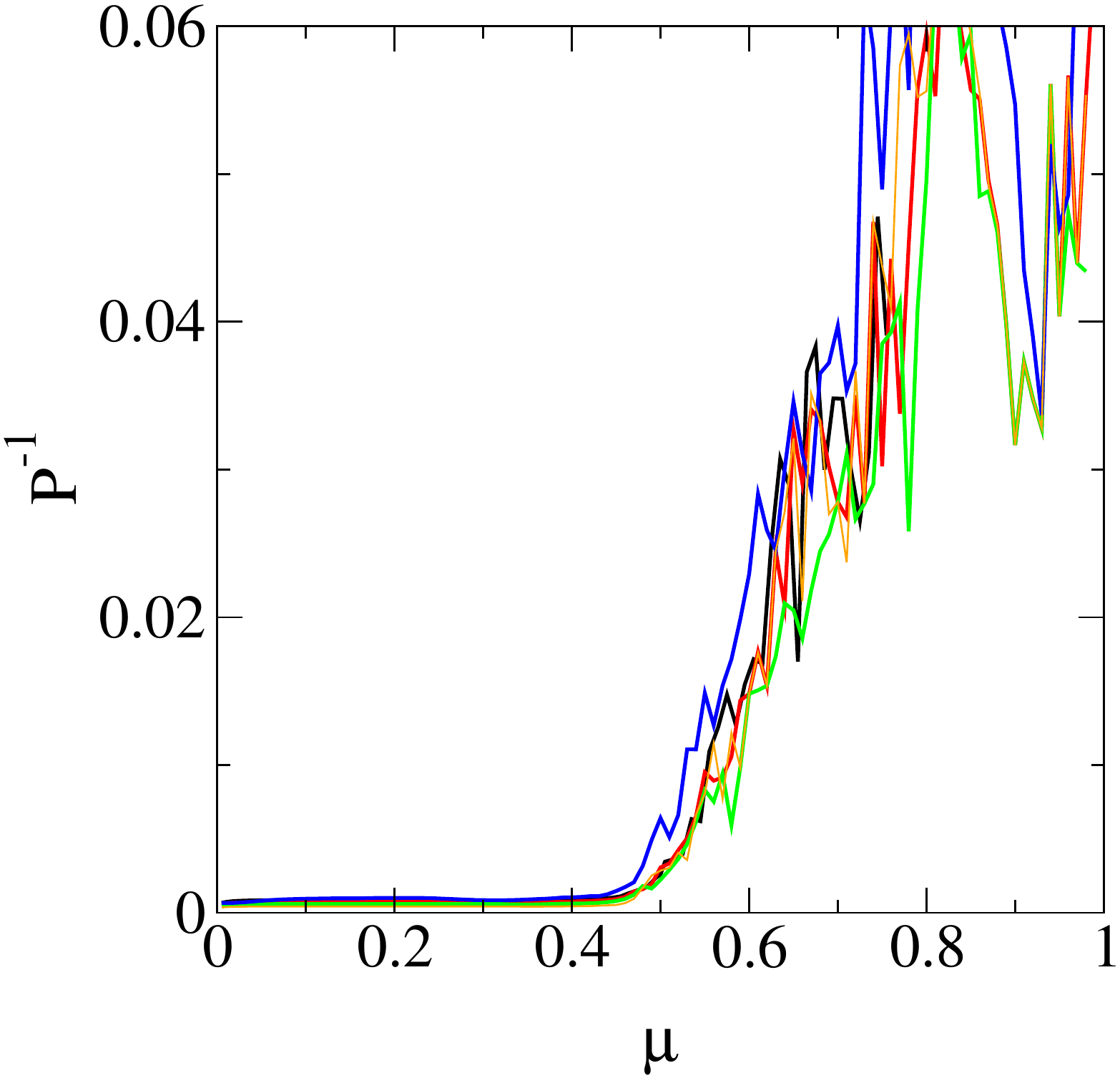}
 \caption{Model A. Inverse of the largest  participation number $P_{max}$ versus $\mu$ for $N=1000,2000,3000,4000,5000$. Here $s=3$, $\lambda=1$ and $K=9$.
}
\label{fig5}
\end{figure}

\begin{figure}[h]
 {\includegraphics[ width=.85\columnwidth]{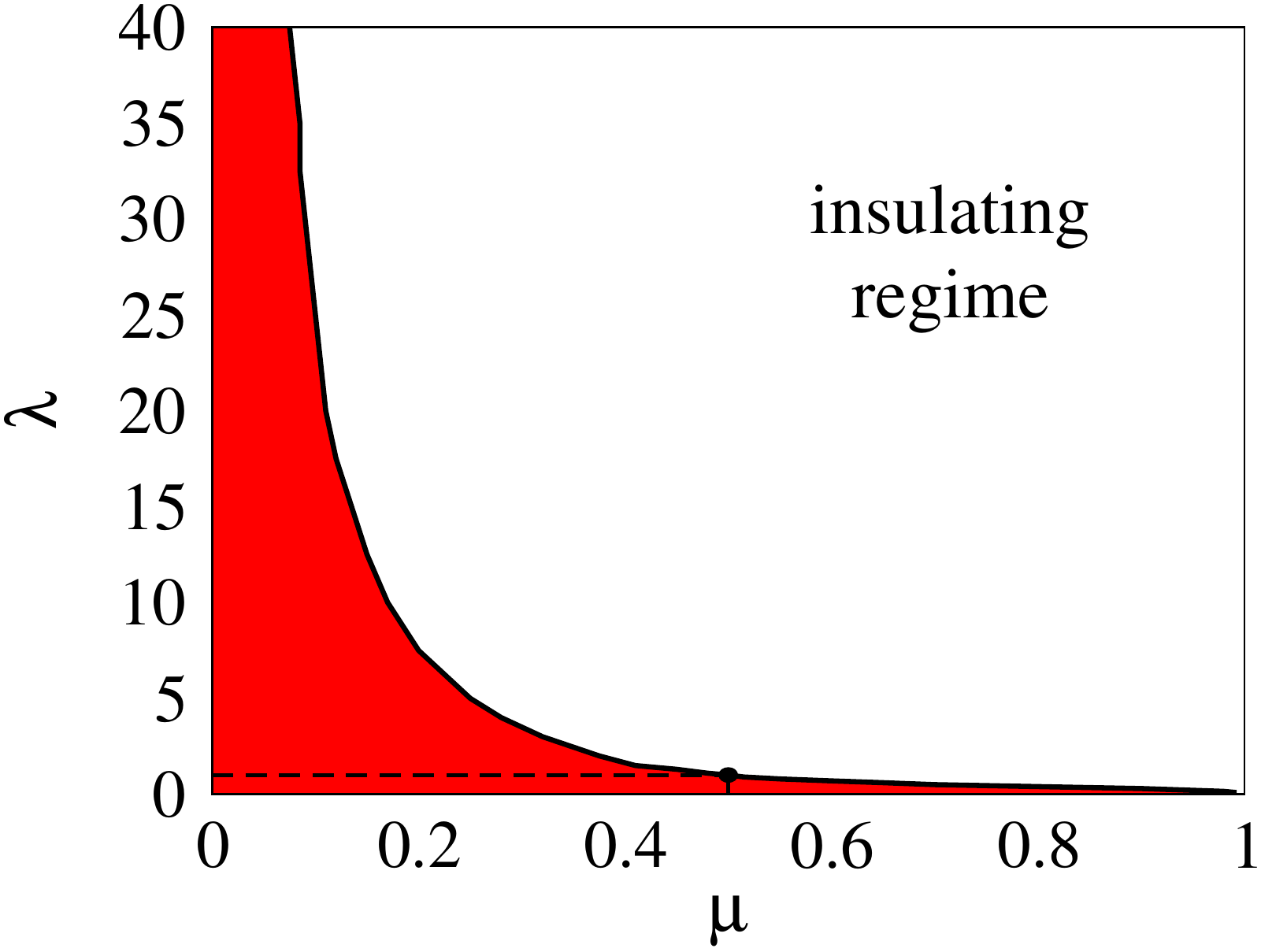}}
 \caption{The phase diagram of model A with potential (\ref{eq:modelA}). The red shaded area corresponds to the metallic regime (extended eigenstates exist). 
Here $s=3$ and $K=9$.
}
\label{fig6}
\end{figure}

We find that $P_{max}^{-1}$ drops down to zero around $\mu = 0.5$ irrespective of the used system size (Fig.\ref{fig5}). These graphs suggest that at $\mu\leq  1/2$ some eigenstates become extended, and so the insulating regime is lost. Calculations for other $\lambda>0$ show how this threshold value of the loss of the insulating regime evolves continuously along the set of parameters $(\mu,\lambda)\in]0,1[\times]0,+\infty[$. The outcome is shown in Fig.\ref{fig6}, where the MIT curve limits the red shaded area in which metallic delocalized states appear.

\begin{figure}[h]
 {\includegraphics[ width=.85\columnwidth]{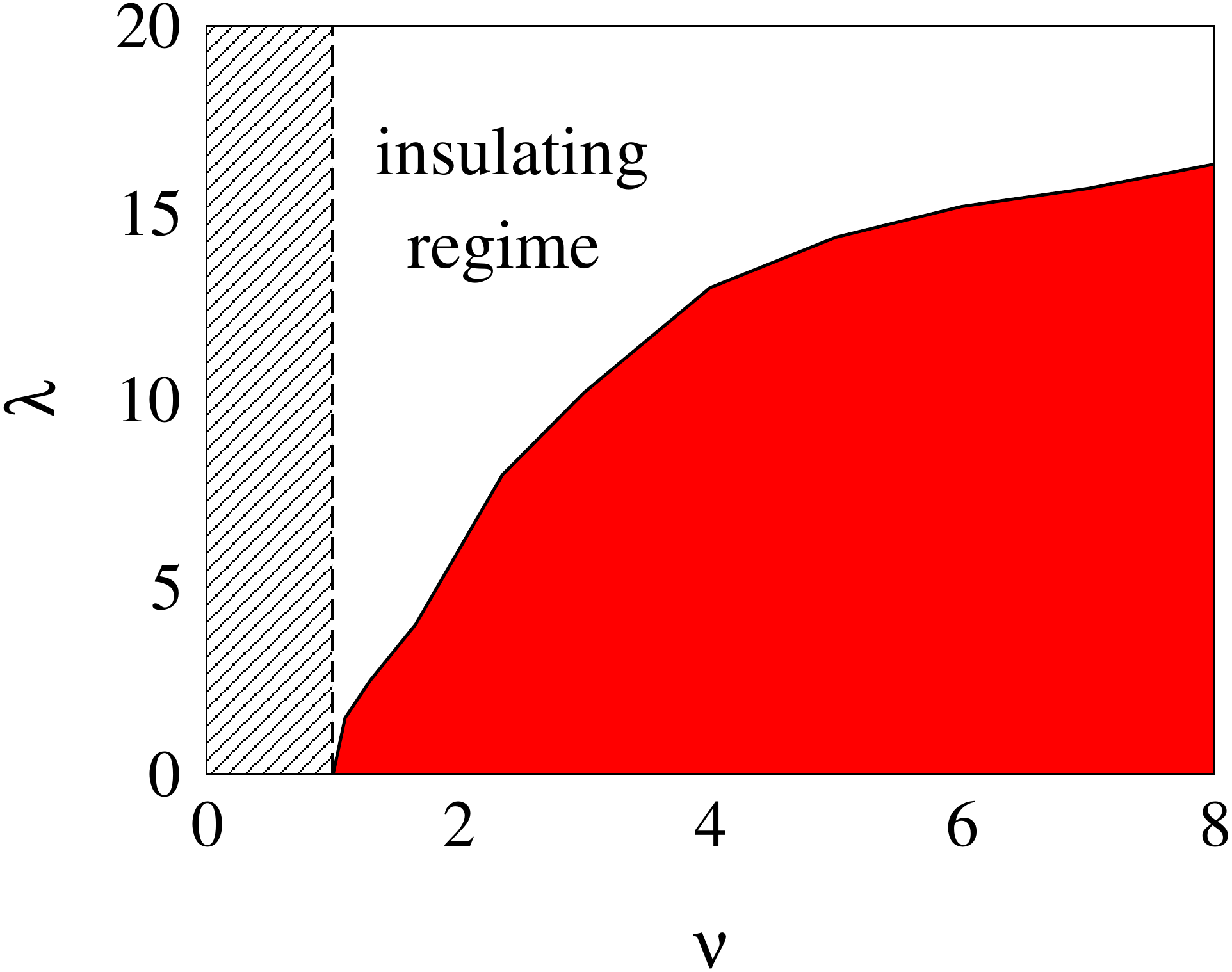}}
 \caption{The phase diagram of model B with potential (\ref{eq:modelB}). The red shaded area corresponds to the metallic regime (extended eigenstates exist). 
Here $s=3$ and $K=9$.
}
\label{fig7}
\end{figure}

A similar MIT arises for model B with the potential (\ref{eq:NPinf}) using an algebraic generating sequence $a_k = 1/k^{\nu}$:
\begin{equation}
\epsilon_{l,K}:=\lambda\sum_{k=1}^K \phi_{k,3}(l) \frac{1}{k^\nu}\ ,\quad \forall l\in\mathbb{Z}\ .
\label{eq:modelB}
\end{equation}
for a real value $\nu\in]1,+\infty[$. Its phase diagram is shown in Fig.\ref{fig7}.

\begin{figure}[!h!]
 \centering
 {\vspace{-3mm}\includegraphics[ width=.85\columnwidth]{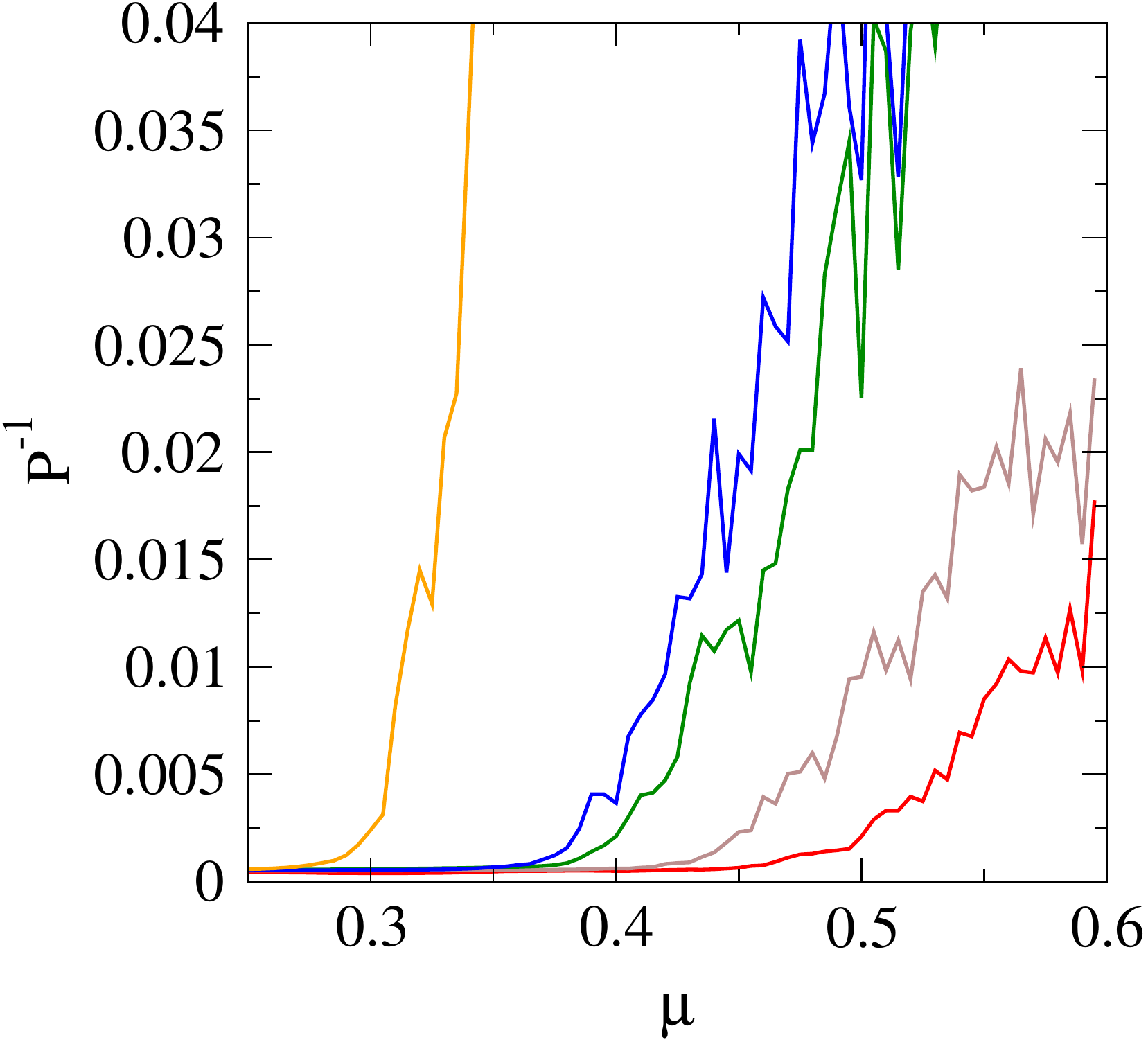}}
 \caption{Model A. Curves of Inverse of the largest Participation Number $P_{max}$ vs $\mu\in]0.25,0.6[$ for $N=5000$ for five evergy sub-intervals. From bottom to top: $0<E<0.31$ (red), $0.31<E<1$ (brown), $1<E<1.58$ (green), $1.58<E<1.92$ (blue), $1.92<E<2.5$ (yellow). Here $s=3$, $\lambda=1$ and $K=9$.}
\label{fig8}
\end{figure}

Returning to model A, 
we compute again for $\lambda=1$ the Participation Number $P$ of eigenstates for different values $\mu\in]0.25,0.6[$. Now we consider energy intervals which correspond to various
sub-bands of the spectrum (see caption in Fig.\ref{fig8} ). In each of these intervals we choose the largest value of $P_{max}$ and plot its inverse as a function of $\mu$ in Fig.\ref{fig8}.

\begin{figure}[h]
 \centering
 {\includegraphics[width=.85\columnwidth]{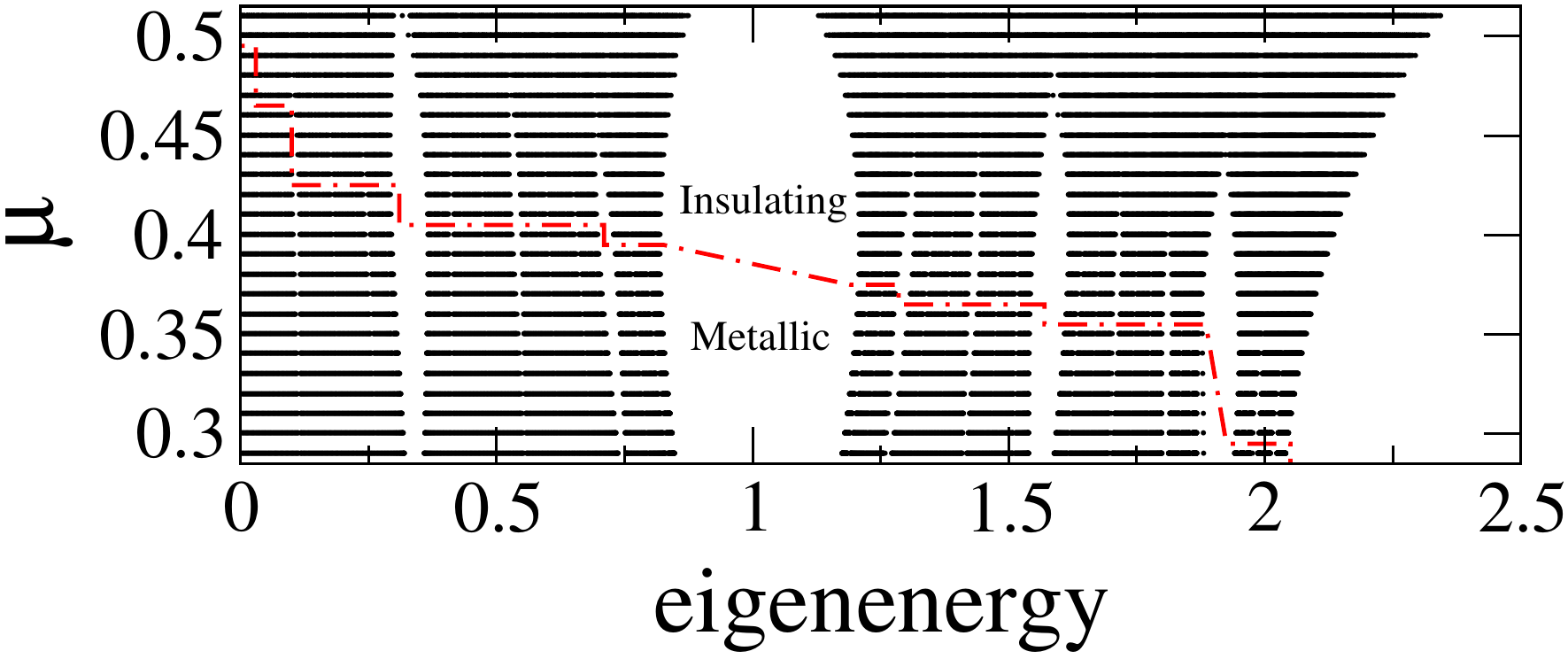}}
 \caption{Model A. A zoom of the eigenenergy spectrum versus $\mu$ from Fig. (\ref{fig:fig4}). The mobility edge is indicated by the thick dashed red line. Here $s=3$, $\lambda=1$ and $K=9$.}
\label{fig9}
\end{figure}

We find that the MIT transition values of $\mu$ and the sharpness of the transition depend on the chosen energy sub-band.
Therefore we observe energy dependent MIT values of $\mu$, i.e. an energy dependent mobility edge Fig.\ref{fig9}.

\begin{figure}[!h]
 \centering
 \includegraphics[width=.44\textwidth]{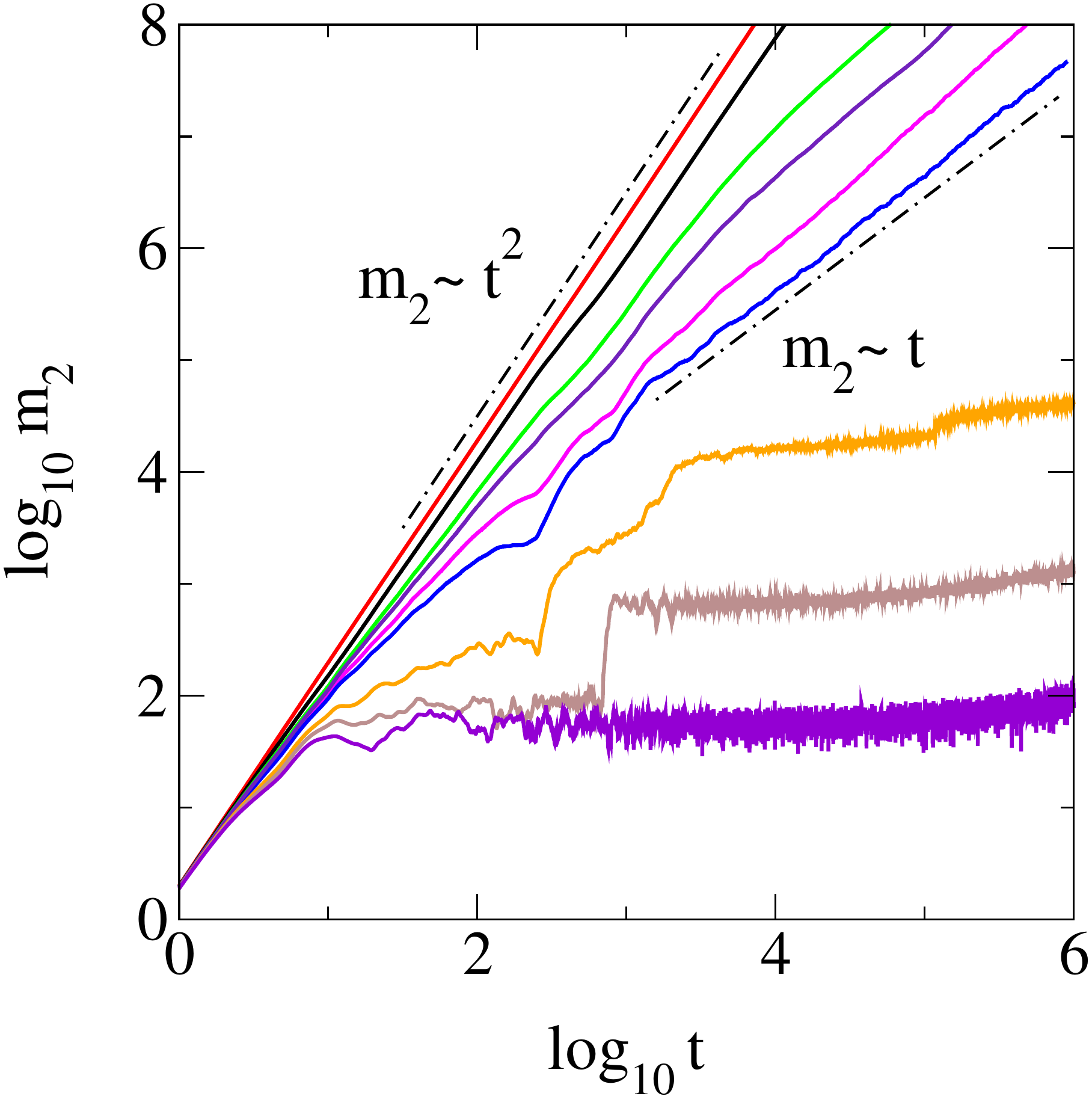}
 \caption{Model A. The second moment $m_2$ for a single site excitation as a function of time in a log-log plot. From top to bottom: $\mu=0.05$ (red), $\mu=0.29$ (black), $\mu=0.4$ (green), $\mu=0.43$ (indaco), $\mu=0.47$ (magenta), $\mu=0.5$ (blue), $\mu=0.6$ (orange), $\mu=0.65$ (brown) and $\mu=0.7$ (purple). The dashed-dotted lines indicate power laws $m_2 \sim t$ (diffusive) and $m_2 \sim t^2$ (ballistic). Here $s=3$,  $\lambda=1$ and $K=9$.
 }
 \label{fig:10}
\end{figure}

Next we study the dynamics for the model A (similar to the AA case).  We compute the time evolution of the second moment $m_2$ for a single site excitation for different 
values $\mu$ of the generating sequence in the interval $]0,1[$ (Fig. \ref{fig:10}). \\

We find that the basic properiets of the AA model are recovered. For small values (around $\mu\sim 0.05-0.1$)  the spreading is  ballistic. An increase of $\mu$ changes the dynamics: an increasingly long transient region
of slower spreading is emerging. At $\mu=0.5$, the long time dynamics is diffusive. For $\mu$ closer to $1$, the wave dynamics show spreading in a finite volume, which will imply localisation.

\section{Summary}

We obtained an iterative construction  of quasiperiodic potentials from sequences of potentials with increasing spatial period. 
At each finite iteration step the eigenstates reflect the properties of the limiting quasiperiodic potential properties up to a controlled maximum system size.
We observe approximate metal-insulator transitions (MIT) at finite iteration steps. We observe mobility edges, at variance to the celebrated Aubry-Andre model. The dynamics near the MIT shows a critical slowing down of the ballistic group velocity in the metallic phase. 
An important open question concerns the existence of suitable choices of the generating sequence (and even different periodic modulations) so that the model has a duality principle. 
In particular it would be interesting to find such a special choice which will reobtain the Aubry-Andre case.

\subsection*{Acknowledgments}

We thank Xiaoquan Yu and  Joshua Bodyfelt for useful discussions.



\begin{thebibliography}{}

\bibitem{anderson1958} P. W. Anderson, Phys. Rev. {\bf 109}, 1492
  (1958).

\bibitem{correlateddisorder} 
F. Izrailev and A. A. Krokhin, Phys. Rev. Lett. {\bf 82}, 4062 (1999).

\bibitem{AA} S. Aubry, G. Andre, Ann. Israel Phys. Soc., vol. 3, Hilger, Bristol, 1980, pp. 133-164
  
\bibitem{grempel82}
D. R. Grempell, S. Fishman and R. Prange, Phys. Rev. Lett. {\bf 49}, 833 (1982).

\bibitem{kohmoto83}
M. Kohmoto, L. P. Kadanoff and C. Tang, Pys. Rev. Lett. {\bf 50}, 1870 (1983).

\bibitem{ostlund83}
S. Ostlund, R. Pandit, D. Rand, H. J. Schellnhuber and E. D. Siggia, Phys. Rev. Lett. {\bf 50}, 1873 (1983).

\bibitem{hiramoto89}
Hisachi Hiramoto and Mahito Kohmoto, Phys. Rev. B {\bf 40}, 8225 (1989).

\bibitem{Flach12} S. Flach, M. Ivanchenko, R. Khomeriki, EPL {\bf 98} (2012) 66002

\bibitem{lahini09}
Y. Lahini, R. Pugatch, F. Pozzi, M. Sorel, R. Morandotti, N. Davidson, and Y. Silberberg, Phys. Rev. Lett.{\bf 103}, 013901 (2009).


\bibitem{roati08}
G. Roati, C. D. Errico, L. Fallani, M. Fatttori, C. Fort, M. Zaccanti, G. Modugno, M. Modugno and M. Inguscio, Nature {\bf 453}, 895 (2008).

\bibitem{larcher09}
M. Larcher, F. Dalfovo and M. Mogudno, Phys. Rev. A {\bf 80}, 053606 (2009).

\bibitem{Av1} A. Avila, Svetlana Jitomirskaya.
Annals of Mathematics {\bf 170} , 303 (2009).

\bibitem{Av2}  S. Ya. Jitomirskaya, I. V. Krasovsky, Math. Res. Lett. {\bf 9}, 413 (2002).  
 
 \bibitem{Avilla} A. Avila, arXiv:0810.2965 (2008).
 
\bibitem{Avilla3} S. Ya.  Jitomirskaya, Ann. of Math. {\bf 150} 1159 (1999). 
  
\bibitem{Avilla2} A. Gordon, S. Ya. Jitomirskaya, Y. Last and B. Simon, Acta Math {\bf 178}, 169 (1997).

\bibitem{Tang} C. Tang and M. Kohmoto, Phys. Rev. B, {\bf 34}, 2041 (1986).

\bibitem{hiramoto88}
H. Hiramoto and S. Abe, J. Phys. Soc. Jpn {\bf 57}, 1365 (1988).

\bibitem{rayanov13}
K. Rayanov, G. Radons and S. Flach, Phys. Rev. E {\bf 88}, 012901 (2013).




 

\end{thebibliography}
\end{document}